\documentclass[12pt,preprint]{aastex}

\def\ltsima{$\; \buildrel < \over \sim \;$}

\def\lsim{\lower.5ex\hbox{\ltsima}}
\def\gtsima{$\; \buildrel > \over \sim \;$}
\def\gsim{\lower.5ex\hbox{\gtsima}}

\begin{document}
\title{Low-Frequency Gravitational Waves from Massive Black Hole
Binaries: Predictions for LISA and Pulsar Timing Arrays}

\author{J. Stuart B. Wyithe\altaffilmark{1} and Abraham
Loeb\altaffilmark{2,3}}

\email{swyithe@isis.ph.unimelb.edu.au; loeb@sns.ias.edu}

\altaffiltext{1}{The University of Melbourne, Parkville, 3010, Australia}

\altaffiltext{2}{Institute for Advanced Study, Princeton, NJ 08540}

\altaffiltext{3}{Guggenheim Fellow; on sabbatical 
leave from the Astronomy Department,
Harvard University}

\begin{abstract}
\noindent 

The coalescence of massive black hole (BH) binaries due to galaxy
mergers provides a primary source of low-frequency gravitational
radiation detectable by pulsar timing measurements and by the
proposed LISA (Laser Interferometry Space Antenna) observatory.  We
compute the expected gravitational radiation signal from sources at
all redshifts by combining the predicted merger rate of galactic halos
with recent measurements of the relation between BH mass, $M_{\rm
bh}$, and the velocity dispersion of its host galaxy, $\sigma$. Our
main findings are as follows: (i) the nHz frequency background is
dominated by BH binaries at redshifts $z\la 2$, and existing limits
from pulsar timing data place tight constraints on the allowed
normalization and power-law slope of the $M_{\rm bh}$--$\sigma$
relation or on the fraction of BH binaries that proceed to
coalescence; (ii) more than half of all discrete mHz massive BH sources
detectable by LISA are likely to originate at redshifts $z\ga 7$;
(iii) the number of LISA sources per unit redshift per year should
drop substantially after reionization as long as BH formation is
triggered by gas cooling in galaxies.  Studies of the highest redshift
sources among the few hundred detectable events per year, will
provide unique information about the physics and history of black hole
growth in galaxies.

\end{abstract}


\section{Introduction}

A future detection of gravitational waves (GWs) will not only test the
theory of General Relativity, but will also provide a precious new
tool in astronomy (Hughes et al.~2001, and references therein). This
tool will be particularly effective for studying compact objects
because the information carried by GWs is very different from that of
electromagnetic radiation. Gravitational radiation is generated by
different physical processes and is often emitted from dense regions
that are otherwise \emph{hidden} from traditional astronomical
observations. A very important distinction, particularly for the study
of the distant universe is that while the usual observable for 
electromagnetic radiation is an energy flux which drops in proportion
to distance squared, the direct observable for gravitational radiation is the
wave amplitude $h$ which drops off only as distance. As a result, the
coalescence of many massive black-hole (BH) binaries ($M_{\rm
bh}>10^3M_\odot$) will be detectable by the planned GW observatory
LISA (Laser Interferometry Space Antenna\footnote{See
http://lisa.jpl.nasa.gov/index.html}) at mHz frequencies out to very
high redshifts (Haehnelt~1994; Hughes et al.~2001).  At the lower
$\mu$Hz frequencies, pulsar timing measurements limit the level of the
cosmic GW background, contributed mainly by the early phase in the
coalescence of massive BH binaries (Kaspi et al.~1994; Thorsett \&
Dewey~1996; Lommen~2002).

Massive BH binaries are among the primary sources for LISA. Such
binaries are produced by mergers of galaxies with pre-existing black
holes, provided that coalescence due to the interaction of the binary
with its surrounding galaxy occurs in less than a Hubble time. 
Hughes~(2002) has shown that LISA observations of the gravitational
radiation during both the inspiral and ring-down phases of massive BH
binary coalescence can be used to determine the source redshift, while
the source position on the sky can be inferred within a fraction of an
orbit of the LISA satellite.
Recently, Seto~(2002) has noted that consideration of the finite
length of the LISA detector leads to significant improvements in
estimations of the accuracy to which the three-dimensional positions
for coalescing BH binaries can be measured.

Black holes are ubiquitous in low redshift galaxies (e.g. Magorrian et
al. 1998). Their masses, $M_{\rm bh}$, as determined from dynamical
measurements (e.g. Merritt \& Ferrarese~2001; Tremaine et al.~2002) or
from reverberation mapping of AGN (e.g. Ferrarese et al.~2001;
Laor~2001), correlate tightly with the velocity dispersion,
$\sigma_\star$, of their host stellar bulge.  However, the abundance
of massive BHs in galaxies at high redshifts is not well known. In
fact, Menou, Haiman \& Narayanan~(2001) have argued that an occupation
fraction as low as 0.01 at redshift 5 can result through subsequent
mergers, in all local massive galaxies containing massive BHs.
Nevertheless, one might expect massive BHs to be ubiquitous at high
redshifts for the following reason. It is commonly thought that
present day BHs are the dormant remnants of the active galactic nuclei
observed at high redshift. Recent observations of this population have
reached out beyond redshift 6 (Fan et al.~2001). The quasars
discovered are very bright, and must be powered by BH masses of $\ga
10^9$M$_\odot$. Recent modeling by Wyithe \& Loeb~(2002) has shown
that the density of these high-redshift quasars as well as lower
redshift populations can be explained through merger activity using
the same relation between host velocity dispersion and BH mass as
observed locally.  The latter assumption is supported by the recent
observation that the relation between velocity dispersion and BH mass
holds in quasars out to redshift 3 (Shields et al.~2002).  This result
suggests that if BHs are not ubiquitous in galaxies at high redshift,
then it must be a coincidence that the active quasar period increases
at high redshift in inverse proportion to the occupation
fraction. Note also that the inferred active period for a BH at
$z\sim6$ is $\sim10^7$yr or $\sim1\%$ of a Hubble time (Wyithe \& Loeb~2002).  
Thus the minimum occupation fraction at $z\sim6$ is $\sim0.01$.

The emission of GWs requires that the binary coalesce in less than a
Hubble time.  What is the mechanism that brings a wide binary system
with an orbital separation $\ga 1$ kpc into the regime where
coalescence is dominated by gravitational radiation? Several
mechanisms have been discussed in the literature, including dynamical
friction on the background stars (e.g. Begelman, Blandford, \& Rees
1980; Rajagopal \& Romani~1995; Milosavljevic \& Merritt~2001;
Yu~2002), and a gas dynamical effect analogous to that thought
responsible for planetary migration (Gould \& Rix~2000;
Armitage \& Natarajan 2002).  Yu~(2002) has followed the expected
evolution of massive central BH binaries in a sample of nearby
early-type galaxies for which high-resolution photometry exists, and
found that the evolution timescale increases significantly when the BH
binary becomes hard before entering the GW regime.  Spherical galaxies
often lead to coalescence timescales that are longer than a Hubble
time, while highly flattened or triaxial galaxies lead to faster
coalescence. Regardless of the physical mechanism, Haehnelt \&
Kauffmann~(2002) argue that if central binaries do not merge in less
than a Hubble time, then three-body interactions resulting from
subsequent mergers will cause ejection and too much scatter in the
$M_{\rm bh}-\sigma_\star$ relation.  Hereafter, we assign an
efficiency $\epsilon_{\rm mrg}$ to the coalescence process. Throughout
this paper we show numerical results under the assumption that BH
binaries formed through the merger of two galaxies proceed quickly to
coalescence ($\epsilon_{\rm mrg}=1$), but point out the relevant
scaling with $\epsilon_{\rm mrg}$ where appropriate.

The number of binary sources depends on a convolution over redshift of
the merger rate of galaxies with the fraction of galaxies that harbor
massive BHs. Several attempts have been made to compute the event rate
(Fukushige, Ebisuzaki \& Makino~1992; Haehnelt~1994; Rajagopal \&
Romani~1995; Haehnelt~1998; Menou, Haiman \& Narayanan~2002; Jaffe \&
Backer~2002), either based on observational counts of quasars and
elliptical galaxies with arbitrary extrapolations to higher redshift,
or using merger tree algorithms based the Press-Schechter
theory. Despite the fact that detectors such as LISA are expected to
detect events at exceedingly high redshift, many calculations have
only considered low ($z<5$) redshift sources.  However, recent
evidence of bright quasar activity at $z\sim 6$ (Fan et al.~2001) and
independent evidence that these quasars reside in massive dark-matter
halos (Barkana \& Loeb~2002), suggests that BHs might be as ubiquitous
at the centers of high-redshift galaxies as they are
today. Furthermore, prior to reionization the ability of gas to cool
inside small halos is mainly limited by the efficiency of gas cooling
rather than by the infall of gas from a heated and photoionized
intergalactic medium, and so massive BHs could form inside much
smaller galaxies than are available after reionization (e.g. Barkana
\& Loeb 2000; Haiman, Abel \& Rees~2001; Thoul \& Weinberg~1996).  As
a result it is possible that a substantial fraction of the events
detectable by LISA have originated prior to reionization.  Our paper
explores this possibility.

Recently, Jaffe \& Backer~(2002) have computed the spectrum of
characteristic strain resulting from massive BH mergers in the
$n$Hz-$\mu$Hz regime using a phenomenological approach for the merger
rate and BH mass function up to some maximum redshift. They
found that the logarithmic spectral slope of the stochastic cosmic
background of gravitational radiation that results from inspiraling
BHs is -2/3, independent of the details of their model for the
merger rate and BH mass function. This provides a numerical
confirmation of the general result derived by Phinney~(2001). They
also found that the amplitude of the spectrum is close to the limit
probed by recent pulsar timing experiments, but is sensitive to the
model assumed for the BH mass function and merger rate. In this paper 
we make a full semi-analytic calculation
of the entire merger rate history starting from the highest redshifts
when the first galaxies are believed to have formed in the universe
(Barkana \& Loeb~2001, and references therein).  We employ the
formalisms of Press \& Schechter~(1974) and Lacey \& Cole~(1993), and
compute the resulting redshift-dependent BH mass function
(including the effect of reionization) using the observed relation
between BH mass and halo circular velocity.  We compare our
results with current limits on the cosmic GW background at low
frequencies, and extend the calculation upwards in frequency to the
LISA band.

Throughout the paper we assume density parameter values of
$\Omega_{m}=0.35$ in matter, $\Omega_{b}=0.052$ in baryons,
$\Omega_\Lambda=0.65$ in a cosmological constant, and a Hubble
constant of $H_0=65~{\rm km\,s^{-1}\,Mpc^{-1}}$ (or equivalently
$h=0.65$). For calculations of the Press-Schechter~(1974) mass
function (with the modification of Sheth \& Tormen~1999) we assume a
primordial power-spectrum with a power-law index $n=1$ and the fitting
formula to the exact transfer function of Cold Dark Matter, given by
Bardeen et al.~(1986).  Unless otherwise noted we adopt a present-day
rms amplitude of $\sigma_8=0.87$ for mass density fluctuations in a
sphere of radius $8h^{-1}$Mpc.

The paper is organized as follows. In \S~\ref{bhev} we describe
calculation of the evolution of the BH mass-function following
reionization. In \S~\ref{mrate}, \S~\ref{edur} and \S~\ref{spec} 
we use the results
of this calculation to predict the merger rate and hence the detection
rate of GW sources by LISA, the distribution of event durations, and the 
resulting spectrum of the
characteristic strain. Finally, we present our conclusions in
\S~\ref{conc}.

\section{Evolution of the Mass-Function of Halos Hosting Massive Black Holes}
\label{bhev}

During hierarchal galaxy formation, an average property of the galaxy
population, such as the density of galaxies evolves in redshift due to
newly collapsing halos, mergers of halos and accretion. In this
section we discuss the calculation of the rate of newly collapsing
halos, and then demonstrate a method to calculate the continuous
evolution of the massive BH mass function.

Let $\frac{dn_{\rm ps}}{dM}(z)$ be the Press-Schechter mass function
(number of halos with mass between $M$ and $M+dM$ per co-moving Mpc)
of dark-matter halos at a redshift $z$. We include the extension of
Sheth \& Tormen~(1999) which improves the agreement with direct
numerical simulations (Jenkins et al.  2001).  The rate of change
(with respect to redshift) of the density of dark matter halos with
masses between $M_1$ and $M_1+dM_1$ is therefore $\frac{d^2n_{\rm
ps}}{dM_1dz}$. Let us also define $\frac{d^2N_{\rm mrg}}{d\Delta
Mdt}\left.\right|_{M}$, the number of mergers per unit time of halos
of mass $\Delta M$ with halos of mass $M$ (forming new halos of mass
$M_1=M+\Delta M$) at redshift $z$ (Lacey \& Cole~1993). 

We consider mergers of halos having masses greater than 
$M_{\rm min}=100M_\odot$. Addition of mass from halos smaller than 
$M_{\rm min}$ is treated as smooth accretion. In what follows this 
contribution to the growth of halos must be included to maintain a 
self-consistent calculation. We therefore find the component of the change 
in the density of halos with masses between $M_1$ and $M_1+dM_1$ that 
is due to accretion rather than to mergers.  This is
given by the overall change in the density of halos with mass $M_1$,
minus the change due to merger of halos to form a new halo of mass
$M_1$, plus the change in density of halos of mass $M_1$ that merge
with other halos (a change that was compensated for in the net
$\frac{d^2n_{\rm ps}}{dM_1dz}$). The net change in density of
halos of mass $M_1$ through accretion is then,
\begin{equation}
\frac{d^2n_{\rm acc}}{dM_1dz} = \frac{d^2n_{\rm
ps}}{dM_1dz}-\int_{M_{\rm min}}^{\frac{M_1}{2}}d\Delta M\left.\frac{d^2N_{\rm
mrg}}{d\Delta Mdt}\right|_{M_1-\Delta M}\frac{dt}{dz}\frac{dn_{\rm
ps}}{d(M_1-\Delta M)} + \int_{M_{\rm min}}^{\infty}d\Delta
M\left.\frac{d^2N_{\rm mrg}}{d\Delta
Mdt}\right|_{M_1}\frac{dt}{dz}\frac{dn_{\rm ps}}{dM_1}.
\end{equation}

We are interested in the evolution of the mass function of dark-matter
halos that contain a massive BH. Prior to reionization, the minimum
mass halo inside of which gas can cool is limited by the cooling rate,
and corresponds to a virial temperature above $T_{\rm min}\sim10^4$K
if the gas cools through transitions in atomic hydrogen, or above
$T_{\rm min}\sim10^{2.4}$K if the gas cools through transitions in
molecular hydrogen (Haiman, Abel \& Rees~2000).  Following
reionization, the IGM is heated to $\sim10^4$K. The Jeans mass is
increased by several orders of magnitude, and numerical simulations
find that gas infall is suppressed in larger halos.  There is some
uncertainty about the precise value of the halo circular velocity
below which gas infall is suppressed (e.g.  Thoul \& Weinberg~1996;
Kitayama \& Ikeuchi~2000; Quinn, Katz \& Efstathiou~1996, Weinberg,
Hernquist \& Katz~1997; Navarro \& Steinmetz~1997). We assume that
following reionization, gas can cool inside newly collapsing halos if
their virial temperatures are higher than $T_{\rm min}=10^5$K. We
denote by $F_{\rm acc}$ the fraction of halos which contain a massive BH
after crossing the cooling threshhold through accretion.
 
Halos forming via mergers are assumed to contain a massive BH
if one of the subunits used to make them already contained a BH.  In
addition, if the merger product has a mass larger than $M(T_{\rm
min})$, then a BH could form out of the gas that cools inside
the merger product. We assume that this happens in a fraction $F_{\rm mrg}$
of cases where neither the initial nor the accreted halo contained a
BH. The mass function of halos containing a massive BH
($\frac{dn_{\rm bh}}{dM}$) therefore evolves according to the
following differential equation:
\begin{eqnarray}
\nonumber
\frac{d^2n_{\rm bh}}{dM_1dz} &=& F_{\rm acc}\frac{d^2n_{\rm acc}}{dM_1dz}\Theta[M_1-M(T_{\rm min})]\\
\nonumber
&+&\int_{M_{\rm min}}^{\frac{M_1}{2}}d\Delta M\left.\frac{d^2N_{\rm mrg}}{d\Delta Mdt}\right|_{M_1-\Delta M}\frac{dt}{dz}\frac{dn_{\rm bh}}{d(M_1-\Delta M)}\Theta[M(T_{\rm min})-M_1]\\
\nonumber
&+& \int_{M_{\rm min}}^{\frac{M_1}{2}}d\Delta M\left.\frac{d^2N_{\rm mrg}}{d\Delta Mdt}\right|_{M_1-\Delta M}\frac{dt}{dz}\left\{\left[1-\left(1-\frac{dn_{\rm bh}/d\Delta M}{dn_{\rm ps}/d\Delta M}\right)\left(1-F_{\rm mrg}\right)\right]\right.\\
\nonumber
&&\left.\times\left(\frac{dn_{\rm ps}}{d(M_1-\Delta M)}-\frac{dn_{\rm bh}}{d(M_1-\Delta M)}\right)+\frac{dn_{\rm bh}}{d(M_1-\Delta M)}\right\}\Theta[M_1-M(T_{\rm min})]\\
&-& \int_{M_{\rm min}}^{\infty}d\Delta M\left.\frac{d^2N_{\rm mrg}}{d\Delta Mdt}\right|_{M_1}\frac{dt}{dz}\frac{dn_{\rm bh}}{dM_1},
\end{eqnarray}
where $\Theta$ is the Heaviside step function.
If $F_{\rm acc}=F_{\rm mrg}=1$, this equation maintains a density of
BHs in halos larger than $M(T_{\rm min})$ that is equal to the
Press-Schechter density.  In the above equation, the first term gives
the change in the density of halos containing BHs due to newly
collapsing halos above the threshold virial temperature. The second and
third terms give the increase due to mergers of the density of halos
of mass $M_1$ that contain a BH. The second term gives the
increase for halos below the threshold temperature due to mergers with
a halo already containing a BH. This term will be important
since the mass corresponding to the minimum temperature increases with
time, and because reionization results in an increase of the minimum
mass halo inside which gas can collapse and cool. The third term gives
the increase in the density of halos with virial temperatures above
the threshold temperature that contain BHs. There are two parts
to this term. If the halo of mass $M_1-\Delta M$ contains a black
hole, then so  does the merger product. However if the halo of mass
$M_1-\Delta M$ does not contain a BH then the merger product
only contains a BH if one of the following two conditions
hold: either the accreted halo (of mass $\Delta M$) already contained
a BH, or if it did not then a BH forms in a fraction
$F_{\rm mrg}$ of merger products.  Finally, the fourth term allows for
the loss in the density of halos of mass $M_1$ that contain black
holes due to mergers forming a larger halo. 
There are two limiting cases; $F_{\rm acc}=F_{\rm mrg}=1$ in which
case all galaxies with masses above the cooling threshhold contain
black-holes, and $F_{\rm acc}=F_{\rm mrg}=0$ in which case the number
of massive BHs drops with time through galaxy mergers. Below we
discuss these cases in more detail.

\begin{figure*}[hptb]
\epsscale{1}
\plotone{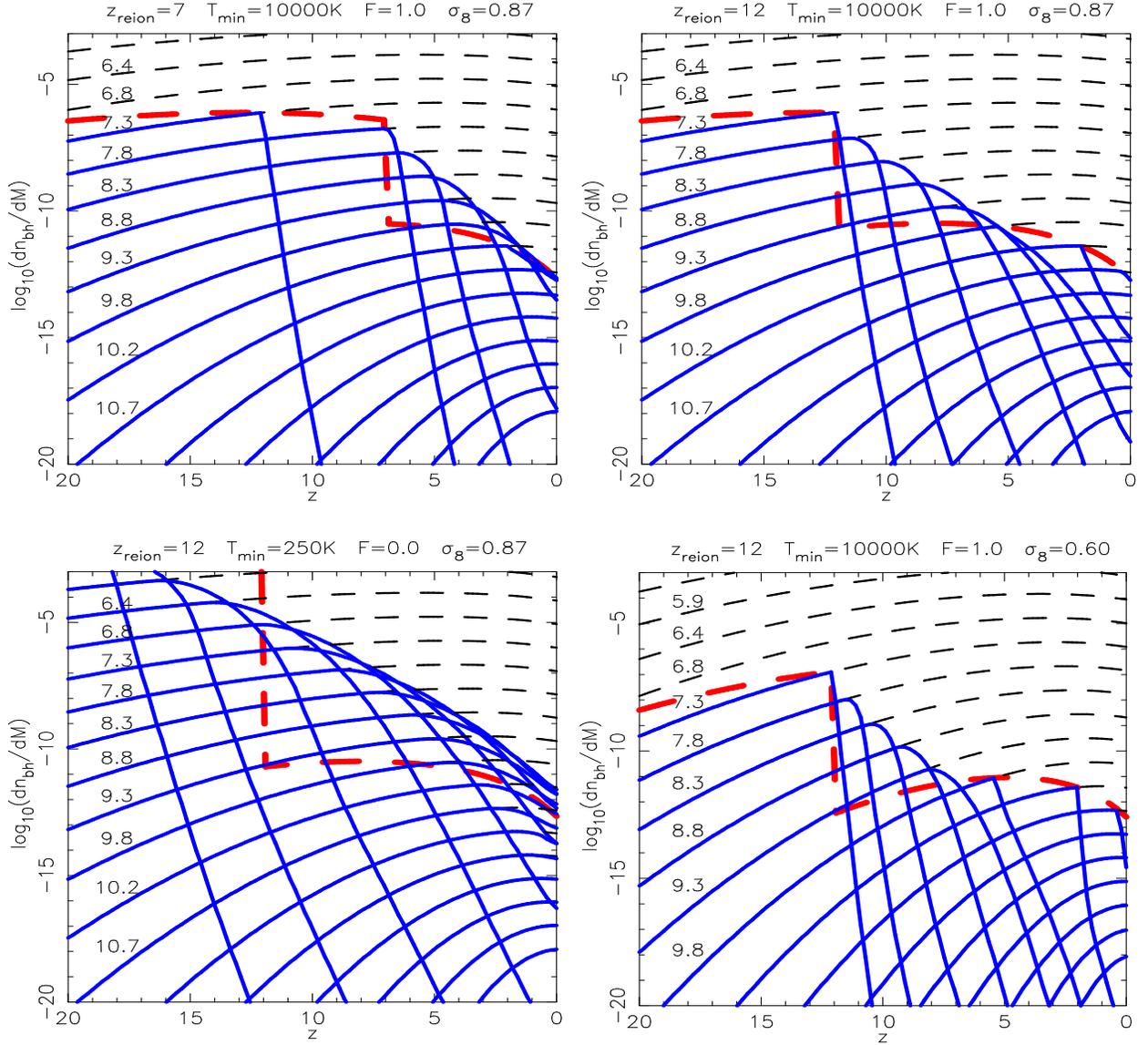}
\caption{\label{fig1} {\footnotesize Solid lines: The mass function of
halos that contain BHs as a function of redshift for various
masses (labeled by $\log_{10}(M/M_\odot)$). Dashed lines: The
Press-Schechter mass function of halos as a function of redshift for
various masses. The panels show cases I-IV whose primary parameters
are listed at the top of each plot. The thick dashed line shows the
density corresponding to $T_{\rm min}$ as a function of redshift.}}
\end{figure*}

We consider 4 examples for the evolution of $\frac{dn_{\rm bh}}{dM_1}$. 
These will be referred to throughout the paper and are labeled $I-IV$. In each 
case the initial values for the mass function of halos containing 
BHs is chosen to be
\begin{equation}
\frac{dn_{\rm bh,ini}}{dM_1}=F_{\rm ini}\frac{dn_{\rm ps}}{dM_1} \Theta[M_1-M(T_{\rm min})].
\end{equation}

\begin{itemize}
\item (I) Begin at $z=20$ with BHs in all halos having 
$T_{\rm vir}>10^4$K, and assume $F_{\rm ini}=F_{\rm acc}=F_{\rm mrg}=1$. 
Reionization is assumed to be at $z_{\rm reion}=7$.

\item (II) Begin at $z=20$ with BHs in all halos having 
$T_{\rm vir}>10^4$K, and assume $F_{\rm ini}=F_{\rm acc}=F_{\rm mrg}=1$. 
Reionization is assumed to be at $z_{\rm reion}=12$.

\item (III) Begin at $z=20$ with BHs in all halos having 
$T_{\rm vir}>10^{2.4}$K, and assume $F_{\rm ini}=1$ and $F_{\rm acc}=F_{\rm mrg}=0$. 

\item (IV) Begin at $z=20$ with BHs in all halos having $T_{\rm
vir}>10^4$K, and assume $F_{\rm ini}=F_{\rm acc}=F_{\rm mrg}=1$.
Reionization is assumed to be at $z_{\rm reion}=12$, but in difference
from case II , we use $\sigma_8=0.6$ rather than the fiducial value
of $\sigma_8=0.87$.
\end{itemize}
Cases I, II and IV refer to situations where cooling in neutral gas is
limited by atomic hydrogen, and in which BH formation is
ongoing. Case III refers to situations where cooling in neutral gas is
limited by molecular hydrogen, and in which the BH population
arises from ``seed BHs'' at $z=20$.

We show the four examples of evolution of $\frac{dn_{\rm bh}}{dM}$ in
Figure~\ref{fig1}. Here the density of halos of mass $M$ that contain
a BH is plotted as a function of redshift (solid lines). The dashed
lines show the corresponding Press-Schechter mass function, and each
pair of curves is labeled by the logarithm of the mass in $M_\odot$ on
the left of the plot. The thick dashed line is the density of halos
with the minimum virial temperature into which gas can collapse, and
has a discontinuity at $z_{\rm reion}$. In cases I and II we see that
following reionization, the density of halos below the critical virial
temperature that contain BHs drops as these halos merge to form more
massive halos.  By redshift 2 ($z_{\rm reion}=7$) and 5 ($z_{\rm
reion}=12$), there are no halos containing BHs left with temperatures
below $T_{\rm min}$.

\subsection{The Black-Hole Mass Function}

The calculation of the BH mass function from the mass-function
of halos that contain BHs requires specification of a relation
between the BH and halo mass.  Ferrarese~(2002) has found
a tight, power-law correlation in nearby galaxies between the mass of a
dark-matter halo and the mass of its central BH. Following
Haehnelt, Natarajan \& Rees~(1998) and Wyithe \& Loeb~(2002) we
generalise to higher redshifts assuming a BH mass that scales
as a powerlaw with circular velocity $M_{\rm bh}\propto v_{\rm
c}^\gamma$. Using the expression for halo mass $M$ as a function of
$v_{\rm c}$ (Barkana \& Loeb~2001) we find
\begin{equation}
\label{bh}
M_{\rm bh} =
10^{12}\epsilon_{0}\left(\frac{M}{10^{12}M_\odot}\right)^{\gamma/3}
\left(\frac{\Omega_{\rm
m}}{\Omega_{\rm m}^{\rm z}}\frac{\Delta_{\rm
c}}{18\pi^2}\right)^{\gamma/6}h^{\gamma/3}(1+z)^{\gamma/2},
\end{equation}
where $\Delta_{\rm c}=18\pi^2+82d-39d^2$, $d=\Omega_{\rm m}^{\rm
z}-1$ and $\Omega_{\rm m}^{\rm z}={\Omega_{\rm
m}(1+z)^3}/[{\Omega_{\rm m}(1+z)^3+\Omega_\Lambda}]$.
In this notation, the relation found by Ferrarese~(2002) for
BHs in the local universe has $\epsilon_0=10^{-5.06}$ and
$\gamma=5.46$. In the remainder of the paper masses with the subscript
$_{\rm bh}$ a subscript refer to BH masses, while those without refer to
halo masses. The merger rates are computed in terms of halo masses,
but many expressions implicitly contain conversion from BH to
halo mass or vice versa using equation~(\ref{bh}).

Simple considerations of the regulation of BH growth by feedback from
quasar activity result in $\gamma=5$ (Haehnelt, Natarajan \&
Rees~1998; Silk \& Rees 1998).  Supporting this argument is the recent
finding of Shields et al.~(2002) that the $M_{\rm bh}-\sigma_\star$
relation derived from QSOs shows no redshift dependence.  The growth
of BHs only through hierarchical mergers imply $\gamma=3$. A value of
$\gamma>3$ is therefore required in order to allow BH growth through
accretion that must accompany quasar activity.  With $\gamma=5$,
theoretical luminosity functions require $\epsilon_0=10^{-5.4}$
(Wyithe \& Loeb~2002), and these are the values used throughout this
work unless specified otherwise.  The mass function of BHs is
therefore
\begin{equation}
\frac{dn_{\rm bh}}{dM_{\rm bh}} = \frac{dn_{\rm
bh}}{dM}\frac{3}{\gamma}\frac{M}{M_{\rm bh}}.
\end{equation}

\begin{figure*}[hptb]
\epsscale{1.}
\plotone{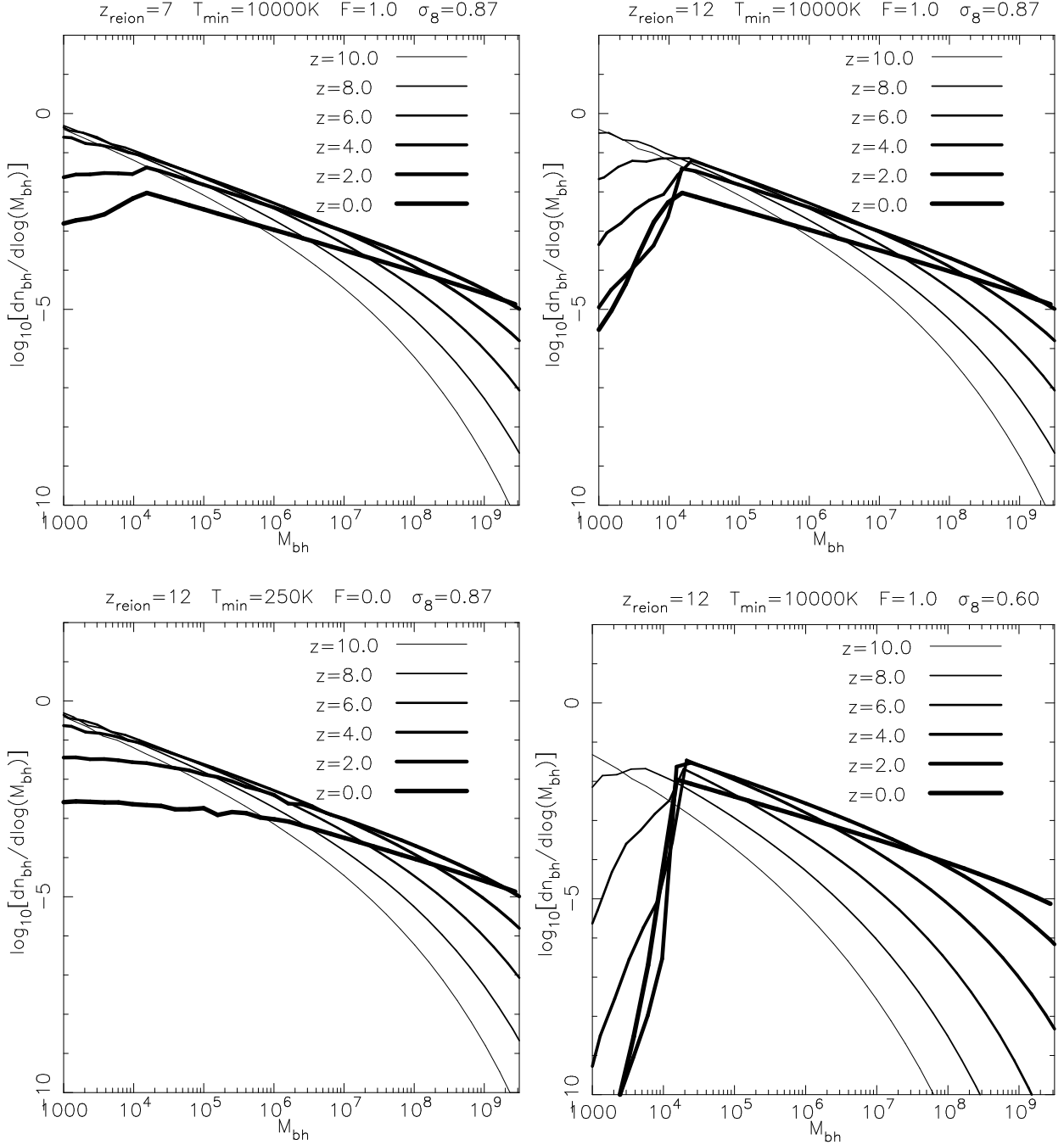}
\caption{\label{fig2} {\footnotesize The mass functions of
massive BHs at various redshifts.  The panels show
cases I-IV whose primary parameters are listed at the top
of each plot.}}
\end{figure*}

The evolution of $\frac{dn_{\rm bh}}{dM_{\rm bh}}$ is shown in each of
cases I-IV in Figure~\ref{fig2}. The mass-function of halos that
contain BHs $\left(\frac{dn_{\rm bh}}{dM}\right)$ is
insensitive to the BH--halo mass relation. However, in the
examples shown, BHs exist in halos as small as
$M\la10^8M_\odot$. Extrapolation of the $M_{\rm bh}-v_{\rm c}$
relation down to these small galaxies is not justified by observations;
moreover, it yields BH masses comparable to those expected from stars
where the formation physics is very different. We therefore only show
the mass function down to $M_{\rm bh}=10^3M_\odot$.  We will show that
the low frequency GW signal is dominated by BHs in galaxies with
$M\ga10^{12}M_\odot$; however BHs in smaller galaxies are important as
seeds for massive BHs in larger galaxies at later times.

The BH mass functions can be compared with those presented in Figure~5
of Kauffmann \& Haehnelt~(2000). The density of the most massive BHs
increases monotonically with time, while in cases I, II and IV there
is little evolution in the density of lower mass BHs before
reionization. However following reionization, there is rapid evolution
at low masses since newly formed low-mass halos do not produce BHs,
and the density of low mass halos that contain BHs drops as they merge
to form larger halos.  Case III shows a different evolution; since
there is no ongoing BH formation the reionization redshift does not
leave any signature.  Case IV with a smaller value of $\sigma_8=0.6$
yields lower BH densities at all redshifts.

\section{Gravitational Waves from Inspiraling Binary Black-Holes}

As described earlier, we assume that in a fraction $\epsilon_{\rm mrg}$
of galaxy mergers the resulting binary BHs evolve rapidly into the
regime where energy loss is dominated by gravitational
radiation. Defining this time as $t_{\rm start}$, the subsequent
evolution of the binary is then described by
equation~(\ref{frequency}). It can be shown (e.g. Shapiro \& Teukolsky
1983) that the evolution in the binaries' (circular) orbital frequency
$f$ with time $t$ obeys the relation
\begin{equation}
\label{frequency}
t(f)-t_{\rm start}=\frac{5}{256}\frac{c^5}{G^{5/3}}\frac{(M_{\rm
bh}+\Delta M_{\rm bh})^{1/3}}{M_{\rm bh}\Delta M_{\rm
bh}}\left(2\pi\right)^{-8/3}\left(f_{\rm
start}^{-8/3}-f^{-8/3}\right),
\end{equation}
where $f_{\rm start}$ is the orbital frequency at time $t_{\rm
start}$. Throughout the paper we assume circular orbits for which the
observed gravitational radiation has a characteristic frequency
\begin{equation}
f_{\rm c}=2\frac{f}{1+z},
\end{equation}
and characteristic strain, averaged over orientations and polarizations
(Thorne~1987) of
\begin{equation}
\label{strain}
h_{\rm c}=8\left(\frac{2}{15}\right)^{1/2}\frac{G^{5/3}M_{\rm bh}\Delta M_{\rm bh}}{c^4R(z)(M_{\rm bh}+\Delta M_{\rm bh})^{1/3}}\left(2\pi f\right)^{2/3},
\end{equation}
where $R(z)$ is the co-moving coordinate distance to redshift $z$. The
choice of $R(z)$ as the distance measure ensures that the energy flux,
which is proportional to $(h_{\rm c}f_{\rm c})^2$, declines in
proportion to the inverse square of the luminosity distance.  As a
binary loses energy, the frequency as well as amplitude of the
gravitational radiation increase with time. Following Hughes~(2002) we 
assume that the inspiral attains its maximum frequency when the two BHs are
separated by 3 Schwarzschild radii, i.e. $f_{\rm
max}=c^3/(2\pi6^{3/2}GM)$Hz for a Schwarzschild BH, and stop our calculation
at this point. Hughes \& Blandford~(2002) find that following a major merger,
the remnant is rarely rotating rapidly unless the mass ratio is $\sim1$.
In this work we neglect the subsequent higher 
frequency radiation from the coalescence and ringdown phases.

\section{The Merger Rate and the Counts of Gravitational Wave Sources}
\label{mrate}

The proposed GW observatory LISA will be able to detect many mergers
of massive BHs out to very high redshifts (Haehnelt~1994; Hughes et
al.~2001). It is therefore interesting to calculate the number of
detectable mergers expected per unit redshift per unit (observed) time
on the sky. Previous attempts at this calculation have used empirical
local rates and extrapolation (e.g. Backer \& Jaffe~2002), or have
used a merger tree algorithm.  Menou, Haiman \& Narayanan~(2001) used
a merger tree algorithm to predict the merger rate out to $z\sim5$,
beyond which they found that the Press-Schechter mass function was no
longer accurately recovered. We have circumvented this problem by
integrating the BH mass function forward in time, while exactly
preserving the Press-Schechter mass function. This allows us to
explore the merger rate at higher redshifts and identify the possible
signature of reionization in GW observations.

\begin{figure*}[hptb]
\epsscale{1.}
\plotone{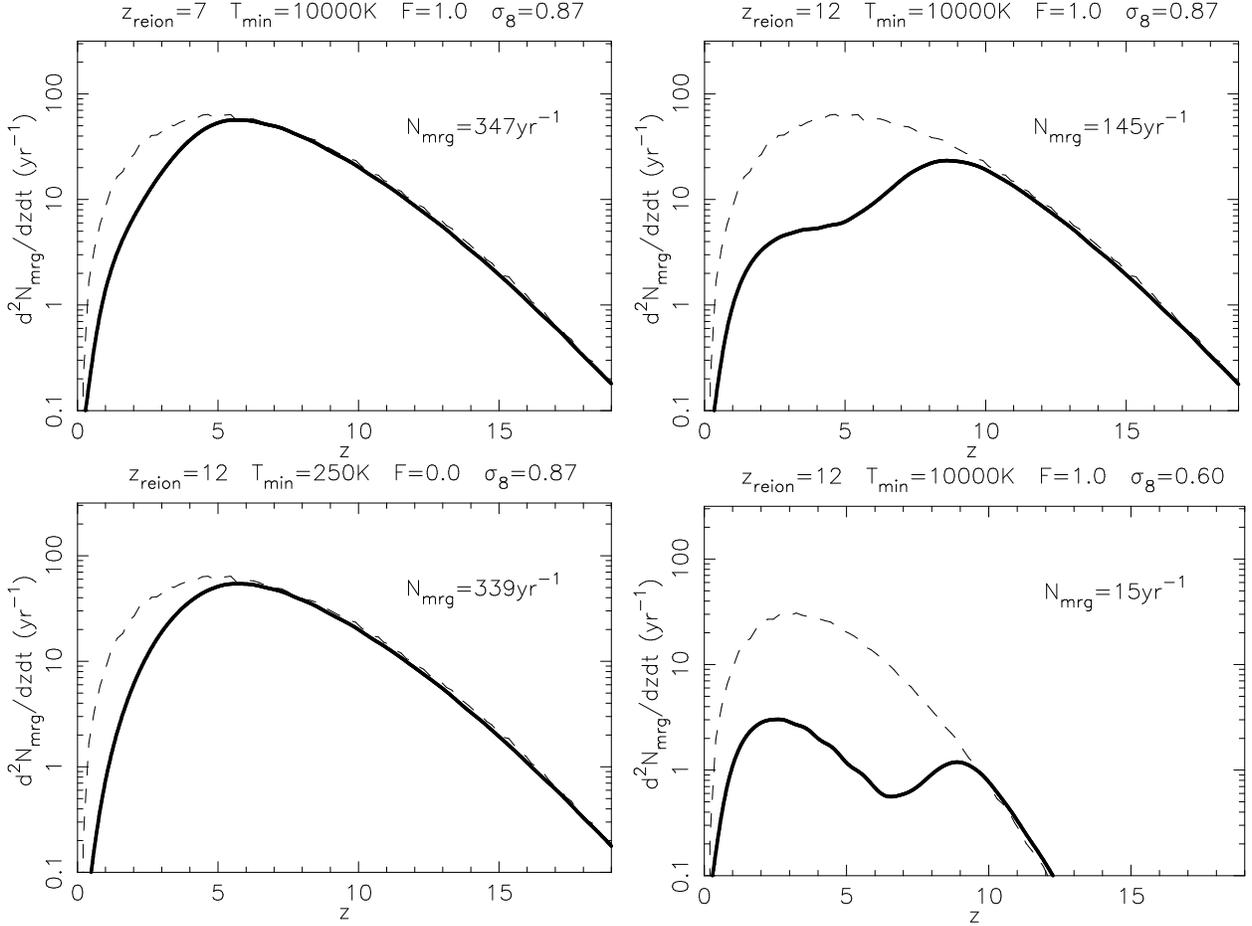}
\caption{\label{fig3} {\footnotesize The rate of GW
events per unit redshift, per observed year.  The panels show
cases I-IV whose primary parameters are listed at the top
of each plot. We assume $\epsilon_0=10^{-5.4}$ and $\gamma=5$. 
For comparison the dashed line shows the event rate
where $T_{\rm min}$ is assumed not to rise following reionization.}}
\end{figure*}

The number of detectable BH mergers per observed year per
unit redshift across the entire sky is
\begin{eqnarray}
\label{mergerrate}
\frac{d^2N_{\rm gw}}{dtdz}&=&\int_0^\infty dM\int_0^{M} d\Delta M\; \Theta(M,\Delta M,f_{\rm c},h_{\rm c},z)\times S(z,M_{\rm bh},\Delta M_{\rm bh})\\
\nonumber
&&\times\left.\frac{d^2N_{\rm mrg}}{d\Delta M dt}\right|_{M} \frac{dn_{\rm bh}}{dM}\left[\frac{dn_{\rm bh}/d\Delta M}{dn_{\rm ps}/d\Delta M}\right]\frac{\epsilon_{\rm mrg}}{1+z}4\pi\frac{d^2V}{dzd\Omega},
\end{eqnarray}
where $V$ is the co-moving volume and $\Omega$ is solid angle on the
sky.  Note that there must be BHs present in both galaxies in order
for gravitational radiation to be produced (this being the origin of
the term in square brackets). The function $\Theta(M,\Delta M,f_{\rm
c},h_{\rm c},z)$ has a value of 1 if the merger produces a detectable
strain in the LISA band, but is zero otherwise. 
In our calculations throughout this paper we set $\epsilon_{\rm mrg}=1$. 
Since $\epsilon_{\rm mrg}$ is our parameterization of the efficiency 
with which dynamical friction brings binary BHs into the GW regime, this 
corresponds to the assumption that all BH binaries coalesce. However, 
more generally, $\frac{d^2N_{\rm gw}}{dtdz}\propto\epsilon_{\rm mrg}$. 

At a frequency of $f_{\rm c}=10^{-3}$Hz, the expected value of the
threshhold sensitivity is $h_{\rm c}\sim10^{-22}$ for a signal to
noise ratio of 5 (needed for confident detection of a binary with
unknown frequency and direction) and one year of observation. We find
that the typical event duration at frequencies between $f_{\rm
c}=10^{-3.5}$Hz and $f_{\rm c}=10^{-2.5}$Hz ranges between a few
months to a year. We therefore assume detectability if the strain
amplitude $h_{\rm c}>10^{-22}$ at an observed frequency of $f_{\rm
c}=10^{-3}$Hz.  Note that a full calculation of the signal to noise ratio of
an observation requires comparison of the coherently folded signal
power (including the evolution in its frequency) to the noise power
accumulated over the bandwidth of the measurement. By working with the
quantity $h_c$ (Thorne~1987), we approximately account for the
detection threshold.

The function $S$ is included in equation~(\ref{mergerrate}) because the 
merger of dark matter halos does not imply the immediate
merger of the stellar systems within. Colpi, Mayer \& Governato~(1999)
have studied the evolution of satellites in virialized halos. They
find that the orbital decay time is
\begin{equation}
t_{\rm decay} = 1.2\left(\frac{r_{\rm vir}}{v_{\rm
c}}\right)\frac{M+\Delta M}{\left[\frac{\Delta
M}{e}\right]\mbox{ln}\left(\frac{M+\Delta M}{\Delta
M}\right)}\epsilon^{0.4},
\end{equation}
where $r_{\rm vir}$ is the initial radius and $\epsilon\sim 0.5$ is
the circularity of the of the orbit (Ghigna et al.~1998).  Note that
even though $\epsilon\sim 0.5$ for the initial orbit of the secondary
galaxy in the dark matter halo, eventually the eccentricity decays
rapidly during the GW driven inspiral of the binary BH system
(e.g. Peters~1964; Hughes \& Blandford~2002; Yu~2002).  The dynamical time may be
written from the expressions in Barkana \& Loeb~(2001) as
\begin{equation}
\frac{r_{\rm vir}}{v_{\rm c}} = 0.10 H^{-1} \left[\frac{\Omega_{\rm m}}{\Omega_{\rm m}^{\rm z}}\frac{\Delta_{\rm c}}{18\pi^2}\right]^{-1/2}h^{-1}(1+z)^{-3/2}
\end{equation}
and amounts to about a tenth of the Hubble time.  In order to have the
satellite sink in less than a Hubble time, the merging galaxies must
be of comparable mass. The dynamical evolution of a massive binary BH
may be disrupted if there is a subsequent major merger before
coalescence of the initial binary. This typically leads to the
ejection of the lightest of the three black-holes (e.g. Haehnelt \&
Kauffmann~2002; Volonteri, Haardt \& Madau~2002). To account for this
phenomenon in a simple way, we assume that at most one coalescence can
occur during the decay time $t_{\rm decay}$ or within a Hubble time,
whichever is the larger.  In mergers where the satellite does not sink
within a Hubble time, we assume no coalescence.  Furthermore,
following Kauffmann \& Haehnelt~(2000) we suppress gas accretion and
therefore BH mergers within galaxies with $v_{\rm c}>600$km/sec,
though this restriction does not significantly affect our results. We
therefore include the function $S(z,M_{\rm bh},\Delta M_{\rm bh})$
which is set to $\left[\mbox{max}\left(\Delta M t_{\rm
decay}\left.\frac{d^2P}{d\Delta Mdt}\right|_M,1 \right)\right]^{-1}$
if the satellite galaxy's orbital decay time $t_{\rm decay}$ is
smaller than the Hubble time, $\left(H_0\sqrt{\Omega_{\rm
m}(1+z)^3+\Omega_\Lambda}\right)^{-1}$, and if both galaxies have
$v_{\rm c}<600$km/sec, but equals zero otherwise.

The quantity $\frac{d^2N_{\rm gw}}{dtdz}$ is plotted in
Figure~\ref{fig3} in the case where $T_{\rm min}=10^4$K at all times
(dashed lines) and in the cases I--IV (solid lines).  The first case
closely resembles the one considered by Menou et al. (2001) in which
they had BHs in all halos (although they did not apply a detection
limit, and hence obtained much higher event rates). The integrated
event rate $N_{\rm mrg}$ is listed in each panel of
Figure~\ref{fig3}. We find that $N_{\rm mrg}$ is sensitive to the
reionization redshift, with values of $N_{\rm mrg}\sim350$yr$^{-1}$
and $N_{\rm mrg}\sim145$yr$^{-1}$ assuming $z_{\rm reion}=7$ and
$z_{\rm reion}=12$ respectively (cases I and II).  As is evident from
the plot, the source counts may probe the reionization epoch out to
large redshifts. There is a drop in the event rate at or soon after
the reionization redshift. This drop arises because even though some
halos below the critical temperature still contain BHs until $z\sim2$
($z_{\rm reion}=7$) and $z\sim5$ ($z_{\rm reion}=12$) respectively,
the event rate scales as the square of the BH occupation
fraction. Similar behavior is seen in case IV, however the smaller
value of $\sigma_8=0.6$ suppresses the growth of structure at early
times relative to our fiducial case of $\sigma_8=0.87$. The resulting
counts are much lower than in cases I-III, $N_{\rm
mrg}\sim15$yr$^{-1}$.  The drop following reionization is much larger
if the suppression of gas infall extends to larger galaxies. For
example, if $T_{\rm min}=2.5\times10^5$K following reionization, the
suppression of the number counts is more than an order of magnitude within
2 redshift units.

Similar results are found with $F_{\rm ini}=F_{\rm acc}=F_{\rm
mrg}=0.01$ ($N_{\rm mrg}\sim340$yr$^{-1}$ and $N_{\rm
mrg}\sim140$yr$^{-1}$ for cases I and II).  Since mergers quickly
result in massive BHs being ubiquitous in larger galaxies even
if the occupation fraction is initially small (Menou, Haiman \&
Narayanan~2001), the number counts
and the corresponding spectra (see subsequent sections) are weakly
dependent on $F_{\rm ini}$, $F_{\rm acc}$ and $F_{\rm mrg}$. We
therefore only consider $F_{\rm ini}=F_{\rm acc}=F_{\rm mrg}=1$ for
cases I, II and IV in the remainder of this paper.

\section{The Distribution of Event Durations of Gravitational Waves from 
Inspiraling BH Binaries}
\label{edur}

In the previous section we calculated the event rate of detectable
gravitational radiation sources for the LISA satellite. However we
would also like to know how long these events will last. To this end
we calculate the distribution of event durations $\Delta t$ for
sources having observed strains between $h_{\rm c}$ and $h_{\rm
c}+\Delta h_{\rm c}$ at $f_{\rm c}=10^{-3}$Hz. We define the event
duration $\Delta t$ as the time taken for the frequency to pass
through a single decade in frequency from $f_{\rm c}=10^{-3.5}$Hz to
$f_{\rm c}=10^{-2.5}$Hz.

\begin{figure*}[hptb]
\epsscale{1.}
\plotone{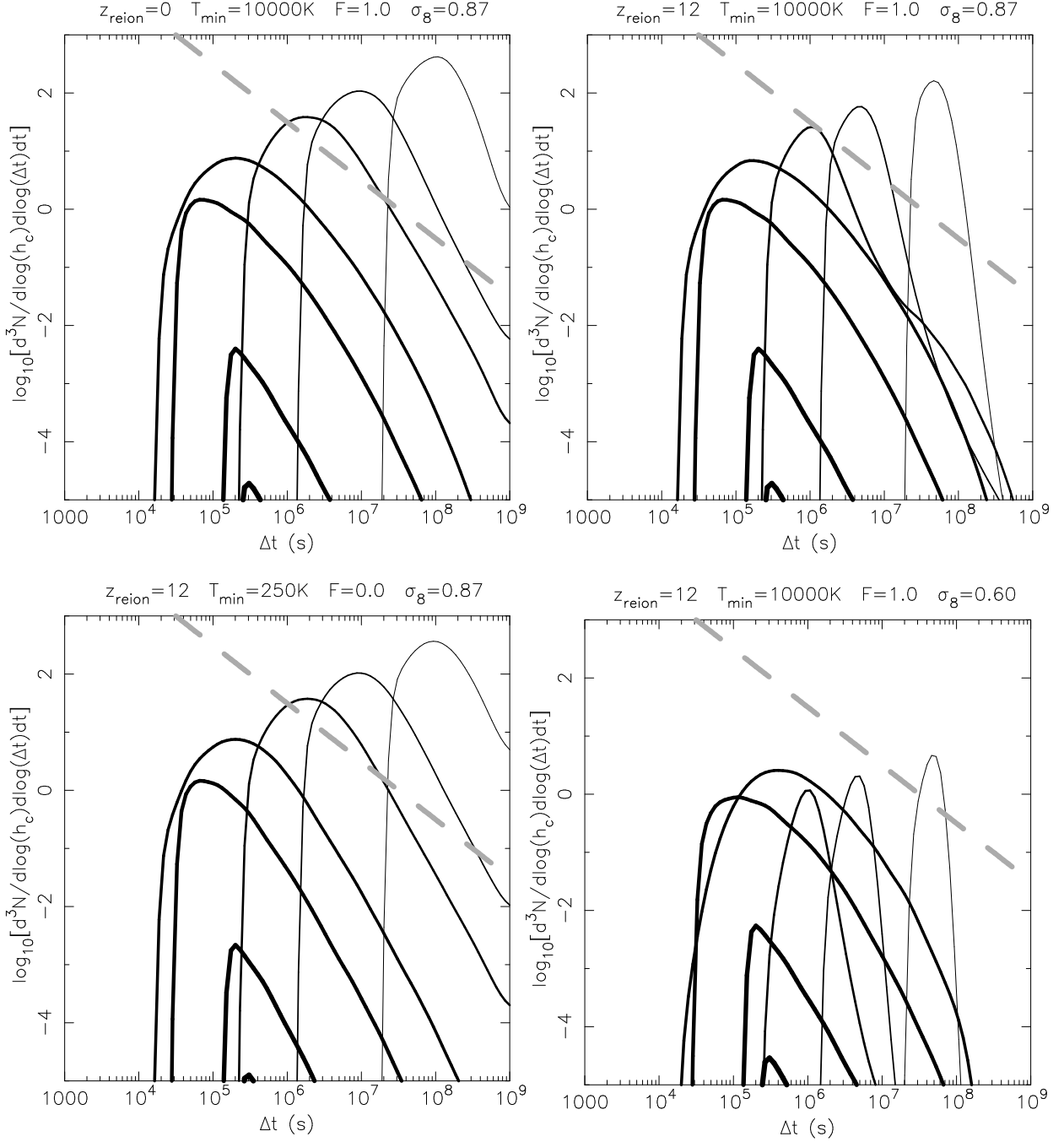}
\caption{\label{fig3a} {\footnotesize Plots of
$\frac{d^3N}{d\log(h_{\rm c})d\log(\Delta t)dt}$ at different values
of strain. Each panel the lines corresponding to $h_{\rm
c}=10^{-23}$, $10^{-22}$, $10^{-21}$, $10^{-20}$, $10^{-19}$ and
$10^{-18}$, from right to left.  Also
shown is the line of $3.15\times10^7\mbox{sec}/\Delta t$ which
approximates the confusion limit for observations with no directional
information.  Curves below this line correspond to radiation from distinct
events. The panels show cases I-IV whose primary parameters are listed
at the top of each plot. We assume $\epsilon_0=10^{-5.4}$ and
$\gamma=5$.}}
\end{figure*}

To calculate the event duration distribution, we assume that a
fraction $\epsilon_{\rm mrg}$ of binary BHs produced through 
galaxy mergers evolve rapidly into the regime where energy loss 
is dominated by
gravitational radiation. We can use the merger rate and the mass
function of halos containing massive BHs together with the BH -- halo
mass relation in equation~(\ref{bh}) to find the observed number of
events of duration $\Delta t$ per unit strain per year. Thus
\begin{eqnarray}
\label{duration}
\nonumber \frac{d^3N}{dh_{\rm c}d\Delta tdt}&=&\int_0^\infty
dM\int_0^M d\Delta M\int_{0}^{\infty}dz \frac{dn_{\rm
bh}}{dM}\left.\frac{d^2N_{\rm mrg}}{d\Delta M dt}\right|_{M}
\left(\frac{dn_{\rm bh}/d\Delta M}{dn_{\rm ps}/d\Delta
M}\right)\times4\pi\frac{\epsilon_{\rm mrg}}{1+z}\frac{d^2V}{dzd\Omega}\\
&&\times\delta\left[K(M_{\rm bh},\Delta M_{\rm bh},z)\right]
\delta\left[T(M_{\rm bh},\Delta M_{\rm bh},z)\right]\times S(z,M_{\rm bh},\Delta M_{\rm bh}),
\end{eqnarray}
where $\delta(x)$ is the Dirac delta function, from equation~(\ref{strain})
\begin{equation}
K(\Delta M_{\rm bh})\equiv h_{\rm
c}-8\left(\frac{2}{15}\right)^{1/2}\frac{G^{5/3}M_{\rm bh}\Delta
M_{\rm bh}}{c^4R(z)(M_{\rm bh}+\Delta M_{\rm bh})^{1/3}}\left(2\pi
f\right)^{2/3},
\end{equation}
and the function $S$, which was introduced in the previous section controls the fraction
of galaxy mergers that can result in BH coalescence within a Hubble time.
From equation~(\ref{frequency}) we find that there is a single redshift corresponding
to each delay for a given combination of merging BH masses, hence 
\begin{equation}
\label{delay}
T(z)\equiv \Delta
t-\frac{40}{256\pi^2}\left(\frac{2}{15}\right)^{0.5}\frac{c}{h_{\rm
c}} \left(10^{-3}~{\rm Hz}\right)^{2/3}\left[\left(10^{-3.5}~{\rm
Hz}\right)^{-8/3} -\left(10^{-2.5}~{\rm
Hz}\right)^{-8/3}\right]\left[R(z)(1+z)^2\right]^{-1}.
\end{equation}

Making the assumption that the timescale for the final inspiral is
short compared with a Hubble time (in the $\epsilon_{\rm mrg}\times S$ of cases
where the satellite sinks to the center of the merger product and the resulting BH 
binary enters the GW regime in less than a Hubble time), we can take the mass-function 
and merger rate terms outside the redshift integral in equation~(\ref{duration}).
If we then integrate over $\delta\left[K(M,\Delta M,z)\right]$ using 
$\int d\Delta M$, and over $\delta\left[T(M,\Delta M,z)\right]$ using 
$\int dz$ we find
\begin{eqnarray}
\label{duration2}
\nonumber
\frac{d^3N}{dh_{\rm c}d\Delta tdt}&=&\\
\nonumber
&&\hspace{-30mm}\int_0^\infty dM \frac{dn_{\rm bh}}{dM}\left.\frac{d^2N_{\rm mrg}}{d\Delta M dt}\right|_M\left[\Delta M_{\rm bh}=K^{-1}(0)\right]\left(\frac{dn_{\rm bh}/d\Delta M}{dn_{\rm ps}/d\Delta M}\right)\times4\pi\frac{\epsilon_{\rm mrg}}{1+z}\frac{d^2V}{dzd\Omega}\times S(z,M_{\rm bh},\Delta M_{\rm bh}) \\
&&\hspace{-30mm}\times  \left(\left|\frac{dK}{d\Delta M_{\rm bh}}\left[\Delta M_{\rm bh}=K^{-1}(0)\right]\right|\frac{d\Delta M_{\rm bh}}{d\Delta M}\right)^{-1} \left(\left|\frac{dT}{dz}\left[z=T^{-1}(0)\right]\right|\right)^{-1}\Theta\left[M_{\rm bh}-K^{-1}(0)\right],
\end{eqnarray}
where we have used the relation 
\begin{equation}
\int dx \delta\left[g(x)\right]=\int dx \sum_i\frac{1}{\left|\frac{dg}{dx}\left[x=g_i^{-1}(0)\right]\right|}\delta\left[x\right].
\end{equation}
The terms are evaluated using the value of $\Delta M_{\rm
bh}=K^{-1}(0)$ found by solving the cubic equation~(\ref{strain}),
which has only one positive real root. The value $z=T^{-1}(0)$ is
found numerically from equation~(\ref{delay}). The Heaviside step
function $\Theta\left[M_{\rm bh}-K^{-1}(0)\right]$ ensures that
mergers are not counted twice. Equation~(\ref{duration2}) has the
following intuitive explanation. For each combination of $M$ and
$\Delta M$ there is a merger rate to which the event rate is directly
related. The derivative $\left(\frac{dK}{d\Delta M_{\rm
bh}}\right)^{-1}$ relates the interval of strain to the interval of
$\Delta M$, thus providing the number density of mergers $\frac{d^2N_{\rm
mrg}}{d\Delta M dt}d\Delta M$ producing events in $dh_{\rm
c}$. Finally the derivative $\left(\frac{dT}{dz}\right)^{-1}$ relates
the interval of event duration to the interval of redshift in which
the source can be located. The volume in which these sources will be
found is then $4\pi\frac{d^2V}{dzd\Omega}dz$.

Curves representing the number of sources per $\log{\Delta t}$ per
$\log{h_{\rm c}}$ per year are plotted in Figure~\ref{fig3a} as
functions of $\Delta t$ for different values of $h_{\rm c}$ in cases
I-IV. Also shown is the confusion limit for observations with no
directional information ($3.15\times10^7\mbox{sec}/\Delta t$).  Note
that this confusion line assumes that the detector has no angular
resolution; in the case of events with locations measured to a
positional accuracy $\Delta \theta$, the confusion line should be
lifted by a factor $(4\pi)/(\pi\Delta\theta^2)$.  Events with values 
of strain corresponding to curves below this line 
will be distinct, while those above the
line will overlap.  Events with lower frequencies come preferentially
from higher redshift. These longer events are more common, which is
consistent with the findings in Figure~\ref{fig3}.  Events with
values of strain $h_{\rm c}\sim10^{-20}-10^{-21}$ and above will be
separable. The figure suggests that in cases I-III around 10 events
per year with $h_{\rm c}\sim10^{-20}$ will be observed, with each
lasting around a few days. At a larger amplitude, $h_{\rm
c}\sim10^{-19}$, about one event per year will be observed lasting
less than a day. Even larger amplitude events from inspirals will be
very rare. In case IV we find that the lower event rate means that
only a few distinct events are expected per year at strains of $h_{\rm
c}\sim10^{-20}$ of more.  Another prominent feature of case IV is the
broadness of the curve for $h_{\rm c}=10^{-20}$, being dominated by
sources producing the peak in number counts near $z\sim4$ (see
Figure~\ref{fig3}).

\section{The Spectrum of the Gravitational Wave Background from Inspiraling 
BH Binaries}
\label{spec}

LISA will be sensitive to the final in-spirals and/or ringdown phases
of most massive BH mergers (Haehnelt~1994; Hughes et al.~2001). 
However, with the exception of the occupation fraction, the
event rate does not directly carry information on properties of the BH
population itself such as the slope and normalization of the BH
mass-function.  Individual events will yield useful information
(Hughes~2002), but we might also expect the number counts of sources
in frequency and strain (i.e. the time averaged number of sources per
unit frequency per unit strain at a single epoch) to provide
constraints on the relation between BH and halo mass.

\begin{figure*}[hptb]
\epsscale{1.}
\plotone{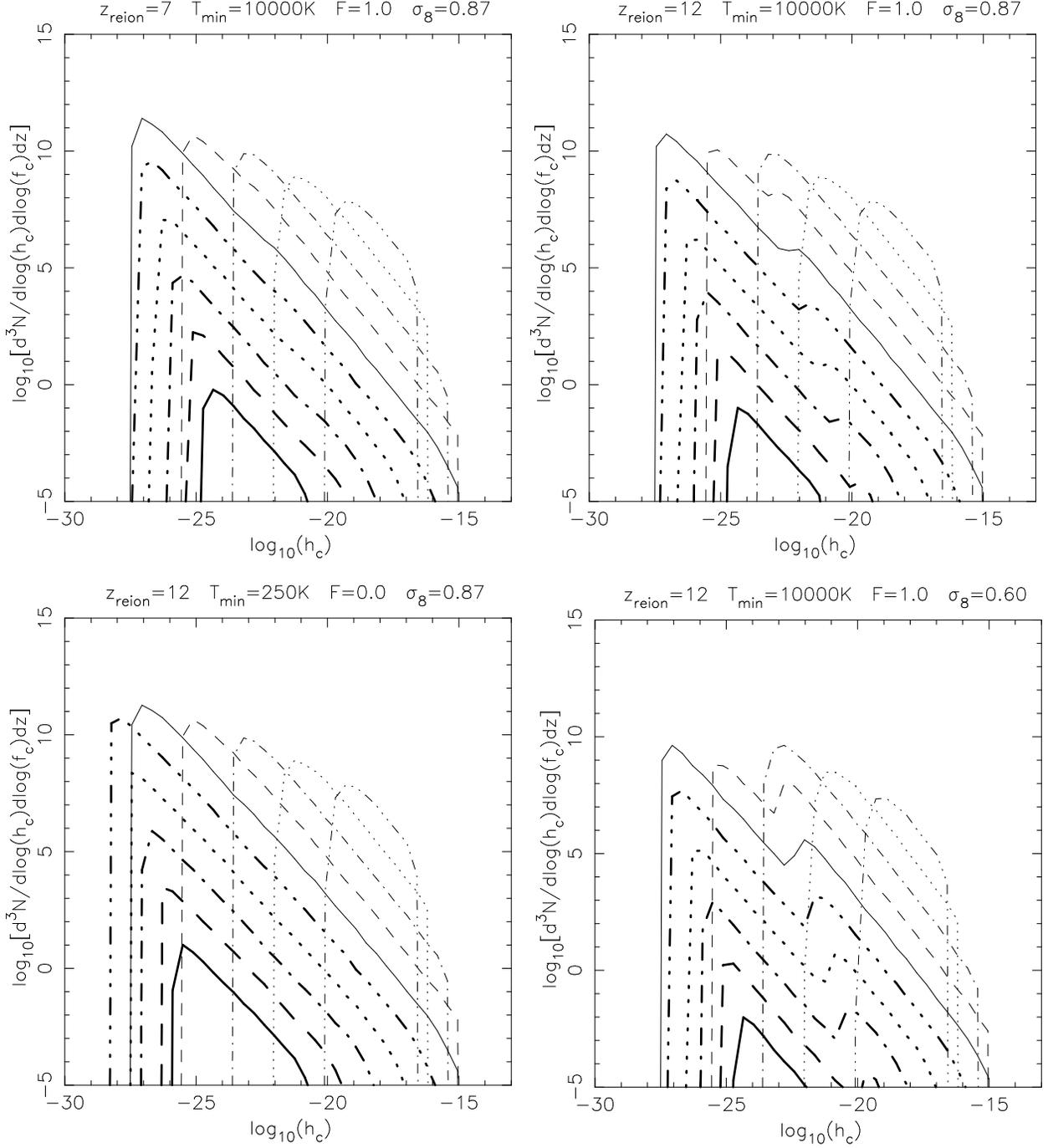}
\caption{\label{fig4} {\footnotesize Plots of
$\frac{d^2N}{d\log(h_{\rm c})d\log(f_{\rm c})}$ at different
frequencies. In each panel the thin lines (from top to bottom)
correspond to $f_{\rm c}=10^{-10}$, $10^{-9}$, $10^{-8}$, $10^{-7}$,
$10^{-6}$, while the thick lines correspond to $10^{-5}$, $10^{-4}$,
$10^{-3}$, $10^{-2}$ and $10^{-1}$Hz.  The panels show
cases I-IV whose primary parameters are listed at the top
of each plot. We assume $\epsilon_0=10^{-5.4}$ and $\gamma=5$.}}
\end{figure*}

In analogy with the previous section we can use the merger rate and
the mass function of halos containing massive BHs together with the BH
- halo mass relation to find the instantaneous, time averaged number
counts that should be observed in a snapshot of the sky per unit
frequency per unit strain per co-moving Mpc,
\begin{eqnarray}
\label{lf}
\nonumber \frac{d^2\Phi}{dh_{\rm c}df_{\rm c}}(z)&=&\int_0^\infty
dM\int_0^M d\Delta M\int_{t_{\rm start}}^{t(z)}dt \frac{dn_{\rm
bh}}{dM}\left.\frac{d^2N_{\rm mrg}}{d\Delta M dt}\right|_{M}\epsilon_{\rm mrg}
\times S(f_{\rm c},z,M_{\rm bh},\Delta M_{\rm bh})\\
&&\hspace{-15mm}\times\left(\frac{dn_{\rm bh}/d\Delta M}{dn_{\rm ps}/d\Delta
M}\right)\delta\left[K(M_{\rm bh},\Delta M_{\rm bh},t-t_{\rm
start})\right]\delta\left[F(M_{\rm bh},\Delta M_{\rm bh},t-t_{\rm
start})\right],
\end{eqnarray}
where from equation~(\ref{frequency})
\begin{equation}
F(t)\equiv f_{\rm c}-\frac{2}{(1+z)}\frac{1}{2\pi}\left[\left(2\pi f_{\rm
start}\right)^{-8/3}-\frac{256}{5}\frac{G^{5/3}}{c^5}\frac{M_{\rm bh}\Delta
M_{\rm bh}}{M_{\rm bh}+\Delta M_{\rm bh}}(t-t_{\rm
start})\right]^{-3/8}.
\end{equation}

Three conditions are imposed on the number counts and are incorporated
in the function $S$.  The first requirement is that the circular
velocity of the merger product be smaller than $600$km/sec. In
\S~\ref{mrate} we assumed that coalescence and the emission of GWs at
$f_{\rm c}=10^{-3}$Hz would only occur where the merging secondary
galaxy sinks before the next comparable merger event 
and in less than a Hubble time. For calculation of
equation~(\ref{lf}) we note that within the GW dominated regime there
are some frequencies and mass combinations for which coalescence does
not occur within a Hubble time. We therefore assume that a BH binary
produced through a galaxy merger will only result in the emission of
gravitational radiation at a frequency $f_{\rm c}$ if the 
binary coalescence as well as the prior sinking of the satellite
occur before the next comparable merger event and within a Hubble
time. Thus the function $S(z,M_{\rm bh},\Delta M_{\rm bh})$ is set to
$\left[\mbox{max}\left(\Delta M \left[t_{\rm decay}+t(f_{\rm
c})-t(f_{\rm max})\right]\left.\frac{d^2P}{d\Delta Mdt}\right|_M,1
\right)\right]^{-1}$ if the satellite galaxy's orbital decay time
$t_{\rm decay}$ plus the inspiral time $[t(f_{\rm c})-t(f_{\rm max})]$
is smaller than the Hubble time, and if both galaxies have $v_{\rm
c}<600$km/sec, but equals zero otherwise.

As before we make the assumption that the timescale for the final inspiral is
short compared with a Hubble time and take the mass-function and
merger rate terms outside the time integral in equation~(\ref{lf}).
If we then integrate over $\delta\left[K(M,\Delta M,t-t_{\rm
start})\right]$ using $\int d\Delta M$, and then over
$\delta\left[F(M,\Delta M,t-t_{\rm start})\right]$ using $\int dt$ we
find
\begin{eqnarray}
\nonumber
\label{spectrum}
\frac{d^2\Phi}{dh_{\rm c}df_{\rm c}}(z)&=&\int_0^\infty dM \frac{dn_{\rm bh}}{dM}\left.\frac{d^2N_{\rm mrg}}{d\Delta M dt}\right|_M\left[\Delta M_{\rm bh}=K^{-1}(0)\right]\left(\frac{dn_{\rm bh}/d\Delta M}{dn_{\rm ps}/d\Delta M}\right) \times S(f_{\rm c},z,M_{\rm bh},\Delta M_{\rm bh})\\
\nonumber
&&\hspace{-20mm}\times  \epsilon_{\rm mrg}\left(\left|\frac{dH}{d\Delta M_{\rm bh}}\left[\Delta M_{\rm bh}=K^{-1}(0)\right]\right|\frac{d\Delta M_{\rm bh}}{d\Delta M}\right)^{-1} \left(\left|\frac{dF}{dt}\left[t=F^{-1}(0)\right]\right|\right)^{-1}\Theta\left[M_{\rm bh}-K^{-1}(0)\right].\\
\end{eqnarray}
The derivative $\frac{dF}{dt}\left[t=F^{-1}(0)\right]$ is easily
computed by substituting $f_{\rm c}$ and $\Delta M_{\rm bh}=K^{-1}(0)$
into $\frac{2}{1+z}\left(\frac{dt}{df}\right)^{-1}$ which is in turn
obtained from equation~(\ref{frequency}). The Heaviside step function
$\Theta\left[M_{\rm bh}-K^{-1}(0)\right]$ ensures that mergers are not
counted twice.

As before, equation~(\ref{spectrum}) has the following intuitive
explanation. For each combination of $M$ and $\Delta M$ the derivative
$\left(\frac{dK}{d\Delta M_{\rm bh}}\right)^{-1}$ relates the interval
of strain to the interval of $\Delta M$, thus providing the number of
mergers per volume per time, $\frac{d^2N_{\rm mrg}}{d\Delta M
dt}d\Delta M$, producing events within $dh_{\rm c}$.  The derivative
$\left(\frac{dF}{dt}\right)^{-1}$ relates the interval of frequency to
the time during which the binary is in that frequency interval. The
number count in a snapshot of the sky is proportional to the number
of mergers during that time.

The number counts of GW sources per logarithm of frequency
per logarithm of strain over the sky is
\begin{equation}
\frac{d^2N}{d\log(h_{\rm c})d\log(f_{\rm c})}=\int_0^\infty dz h_{\rm c}f_{\rm c}\frac{d^2\Phi}{dh_{\rm c}df_{\rm c}}(z)4\pi \frac{d^2V}{d\Omega dz}.
\end{equation}
This quantity is plotted in Figure~\ref{fig4} as a function of $h_{\rm
c}$ for different values of $f_{\rm c}$ and for cases I-IV.  As
expected, the number counts increase toward low frequencies since the
inspiral lifetime scales as $t\propto f^{-8/3}$. However at all
frequencies the number counts are cut off at both high and low
amplitudes. The reasons are as follows. First, the cutoff at large
$h_{\rm c}$ is caused by the fact that the largest amplitude waves are
produced at the final phase of the inspirals. Since the inspiral has a
maximum frequency (Hughes~2002), reached when the binary separation becomes
comparable to the Schwarzschild radius, each binary also has a
corresponding maximum amplitude. This effect plays a significant role
in the characteristic strain spectrum in the frequency range
applicable to LISA, a point to which we will return in the following
section, \S~\ref{strainspec}.  Second, the cutoff at low amplitude is
caused by the fact that at a fixed frequency, $h_{\rm c}\propto M_{\rm
bh}\Delta M_{\rm bh}(M_{\rm bh}+\Delta M_{\rm bh})^{-1/3}$. Since both
$M_{\rm bh}$ and $\Delta M_{\rm bh}$ have a minimum value $M_{\rm
bh,min}$, the strain must be larger than a value proportional to
$M_{\rm bh,min}^2(2M_{\rm bh,min})^{-1/3}$. 

\subsection{The Characteristic Strain Spectrum}
\label{strainspec}

The \emph{characteristic strain spectrum}, $h_{\rm spec}[f_{\rm c}]$
(e.g. Phinney~2001; Jaffe \& Backer~2002), describes the spatial
spectrum of the stochastic background of gravitational radiation.  The
characteristic strain spectrum is related to the strain power-spectrum
$S_{\rm h}$ through
\begin{equation}
\label{hspec}
h_{\rm spec}(f_{\rm c})=\sqrt{f_{\rm c}S_{\rm h}(f_{\rm c})}.
\end{equation}
where $S_{\rm h}(f_{\rm c})$ is calculated from
\begin{equation}
\label{hspec2}
S_{\rm h}(f_{\rm c})=\int_0^\infty dh_{\rm c} \int_0^\infty dz\,h_{\rm c}^2\frac{d^2\Phi}{dh_{\rm c}df_{\rm c}}(z)4\pi \frac{d^2V}{d\Omega dz}.
\end{equation}
The contribution to the characteristic strain spectrum from sources in
a redshift interval can be calculated by integrating between the
desired limits.

\begin{figure*}[hptb]
\epsscale{1.}
\plotone{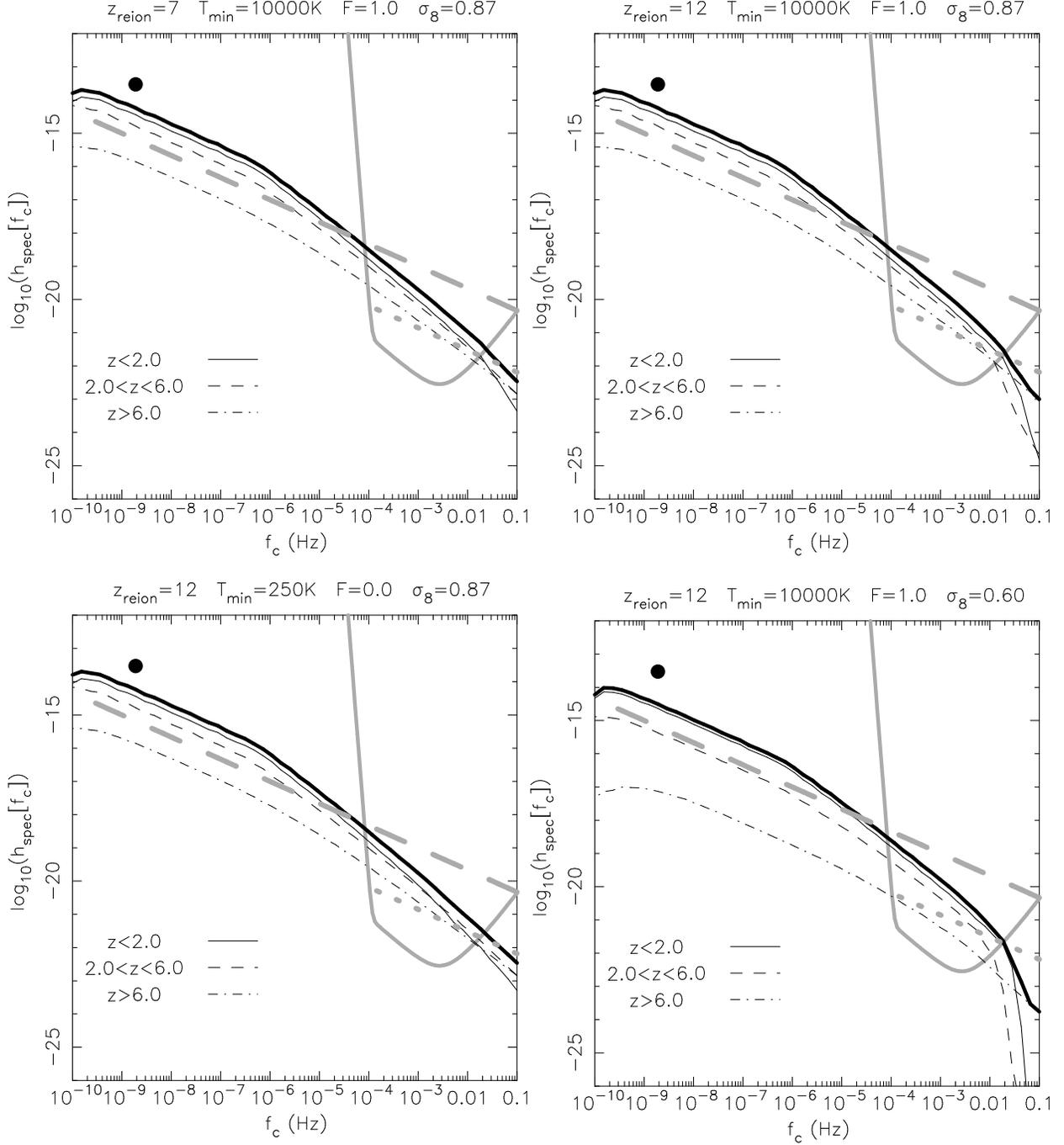}
\caption{\label{fig5} {\footnotesize The characteristic strain
spectrum.  In each panel the thin lines in order from bottom to top
show the spectrum in different redshift intervals and the thick dark
solid line is the total strain spectrum. The thick dashed line shows
the characteristic spectrum $h_{\rm spec}=10^{-21}f_{\rm c}^{-2/3}$
found by Jaffe \& Backer~(2002), and the thick dotted line the
estimated stellar binary foreground from Phinney~(2001).  The thick
grey solid line in the lower right corner of each panel shows the
expected instrumental noise for LISA.  The dot shows the current limit
from pulsar timing. The panels show cases I-IV whose primary
parameters are listed at the top of each plot.  We assume
$\epsilon_0=10^{-5.4}$ and $\gamma=5$.}}
\end{figure*}

We have plotted $h_{\rm spec}(f_{\rm c})$ in Figure~\ref{fig5} for
cases I-IV.  The thin lines from bottom to top show $h_{\rm
spec}(f_{\rm c})$ in the redshift intervals $z>6$, $2<z<6$, and
$z<2$. The solid thick lines show the total $h_{\rm spec}(f_{\rm
c})$. The characteristic strain spectrum is dominated by sources
having $z<2$. The thick dashed line shows the spectrum deduced by
Jaffe \& Backer~(2002). They found $h_{\rm spec}(f_{\rm
c})\sim10^{-21}f_{c}^{-2/3}$ for the region accessible to pulsar
timing experiments.  Figure~\ref{fig5} 
shows their result at nHz frequencies and extrapolates it to the
higher frequencies probed by LISA for comparison.

At frequencies between $10^{-9}$Hz and $10^{-6}$Hz, the characteristic
strain spectrum has a logarithmic slope of -2/3. This slope results
from the dependence of energy loss on frequency as explained by
Phinney~(2001).  At the higher frequencies relevant for LISA, we find
departures from the -2/3 slope. The steepening of $h_{\rm spec}(f_{\rm
c})$ is due to the loss of power at high frequencies associated with
the maximum frequency during inspiral. At frequencies below
$10^{-9}$Hz there is loss of power arising because many very low
frequency sources do not coalesce within a Hubble time.  The thick
grey lines in the lower right corner of each panel in
Figure~\ref{fig5} show the instrumental noise equivalent of $h_{\rm
spec}$ computed from the instrumental noise spectral amplitude $S_{\rm
h}^{\rm inst}(f_{\rm c})$. The curves are from a fit (Hughes 2002) to
calculations of $S_{\rm h}^{\rm inst}(f_{\rm c})$ (Folkner 1998).  The
amplitude of the characteristic strain spectrum should be detectable
by LISA over the frequency range $\sim10^{-4}$--$10^{-2}$Hz. However
the low numbers of events will make the power-spectrum noisy in this
regime. In general the binaries can be resolved provided less than
20\% of the frequency bins are filled at any one time. As we showed in
\S~\ref{edur}, most events are of sufficiently short duration to allow
each to be resolved separately.  The white-dwarf/white-dwarf binary
foreground was estimated by Phinney~(2001) and is plotted as the thick
dotted lines [cut off below $10^{-4}$Hz since low frequency binaries
are not expected to contribute significant gravitational waves (Hils,
Bender, \& Webbink 1991).  Through most of the LISA band, this
foreground is more than an order of magnitude below the expected
background due to coalescing massive BH binaries.

We also show in Figure~\ref{fig5} the current upper limits on the GW
background (large dots). The most recent limits are quoted by
Lommen~(2002). They find a limit on the fractional energy density in
gravitational waves per logarithm of frequency at $f_{\rm
c}=1.9\times10^{-9}$Hz is $\Omega_{\rm g}h^{2}<2\times10^{-9}$.  We
then find a corresponding limit\footnote{ This value is an order of
magnitude larger than that plotted in Figure~8 of Jaffe \&
Backer~(2002). Use of the value in their paper leads to even tighter
limits on the parameters $\epsilon_0$, $\gamma$ and $\epsilon_{\rm
mrg}$} on the characteristic strain spectrum of $h_{\rm
spec}(1.9\times10^{-9}$Hz$)<3\times10^{-14}$. The limits are close to
the theoretical spectrum in cases I-III. We will return to this
important and unexpected result below.

Our results in cases I--III for the amplitude of the characteristic
strain spectrum are different from the findings of Jaffe \&
Backer~(2002) in the regime accessible to pulsar timing. In particular
we find that the models predict $h_{\rm spec}(1.9\times 10^{-9}$Hz$)$
that is an order of magnitude higher than predicted by Jaffe \&
Backer~(2002), and which is close to the limits of detection by the
pulsar timing measurements. There are several possible causes for this
difference. Jaffe \& Backer~(2002) use a power-law extrapolation of an
estimate of the spheroid merger rate to higher redshifts, while in
fact galaxies were made of smaller sub-units that had a higher merger
probability at earlier cosmic times. Also, they use a phenomenological
prescription for the merger rate that may substantially underestimate
the actual merger rate for short duty cycle events.  Furthermore, they
assume the same merger rate for all galaxies, and do not assign a
relation between black-hole and galaxy mass. In essence, we have
assumed BHs to be associated with dark-matter halos, for which the
merger history is understood rather than use an estimate of the
average spheroid merger rate. The two assumptions lead to different
results because the total merger rate of galaxies above the cooling
threshhold as computed from the Press-Schechter formalism is higher
than estimates of the observed spheroid merger rate. We find that if
we artificially force equality between the merger rates computed in
our work and those used in Jaffe \& Backer~(2002), we obtain a
similar spectrum well below current limits.

The results should also be dependent on the BH -- halo mass relation. The
amplitude of $h_{\rm spec}$ is proportional to $\epsilon_0^{5/3}$ [see
equation~(\ref{strain}), (\ref{hspec}) and (\ref{hspec2})].
As a result, a lower value of $\epsilon_0$ will result in $h_{\rm
spec}$ having a lower amplitude.  Moreover, the production of power at
very low frequencies is dominated by very large BHs, and so $h_{\rm
spec}$ should be sensitive to $\gamma$, with larger $\gamma$ values
resulting in larger amplitudes. We note that the log-normal
distribution employed by Jaffe \& Backer~(2002) is steeper at the
highest masses than our BH mass function computed from the
Press-Schechter mass function and the $M_{\rm bh}-M$ relation, which
might further contribute to our differing results. We will return to
the dependence on the BH -- halo mass relation in the following
section (\S~\ref{msiglim}).

Our simulation of Case IV yields a value of $h_{\rm spec}(1.9\times
10^{-9}$Hz$)$ that is an order of magnitude below current limits, 
and which is more consistent with the findings of Jaffe \&
Backer~(2002). Thus, the power in GWs is quite sensitive to the value of
$\sigma_8$. Indeed $h_{\rm spec}$ is significantly more sensitive to
$\sigma_8$ than to the details of the merger history provided that the
history results in BHs being ubiquitous in massive galaxies in
the local universe.

We find that the characteristic strain spectrum at low frequencies is
dominated by galaxies having masses larger than $\sim10^{12}M_\odot$
merging at redshifts below the peak in quasar activity ($z\la2$). This
is particularly true of case IV.  Theoretical studies of the quasar
luminosity function find that the lack of cold gas at low redshifts
explains the drop in quasar activity in massive galaxies\footnote{It 
is worth noting that the fiducial values of
$\epsilon_0$ and $\gamma$ used in this work are taken from results of the 
theoretical quasar luminosity function described in Wyithe \& Loeb~(2002)
which over-predicts the number of bright quasars at low redshift. However 
this approach does not over-predict the GW back-ground by predicting the
existence of to many very massive BHs because similar values of $\epsilon_0$ 
and $\gamma$ have been measured by Ferrarese~(2000) in nearby galaxies.} 
(Kauffmann \& Haehnelt~2000).  If as suggested (e.g. Gould \& Rix~2000) gas that is
driven to the center of the galaxy during a merger is the mechanism
that brings the BH binary into the GW dominated regime, then we might
expect galaxy mergers to be inefficient ($\epsilon_{\rm mrg}\ll1$) at
producing gravitational radiation at low redshifts. If $\epsilon_{\rm
mrg}\ll1$, then we expect the amplitude of $h_{\rm spec}$ which is
proportional to $\sqrt{\epsilon_{\rm mrg}}$, to be lower than
predicted. Our results from cases I-III imply that limits on the
efficiency can already be placed for some values of $\gamma$ and
$\epsilon_0$ using limits on $h_{\rm spec}$ from pulsar timing
experiments (\S~\ref{msiglim}). If $\epsilon_{\rm mrg}$ is found to
have a value of order unity through the detection of the background
by future pulsar timing experiments or by the LISA satellite, then
this would strengthen the case for similar values at yet higher
redshifts. The number counts of mergers detected by LISA will
therefore directly determine the BH occupation fraction at high
redshift.

\subsection{Limits on $\epsilon_0$ and $\gamma$}
\label{msiglim}

\begin{figure*}[hptb]
\epsscale{1.}
\plotone{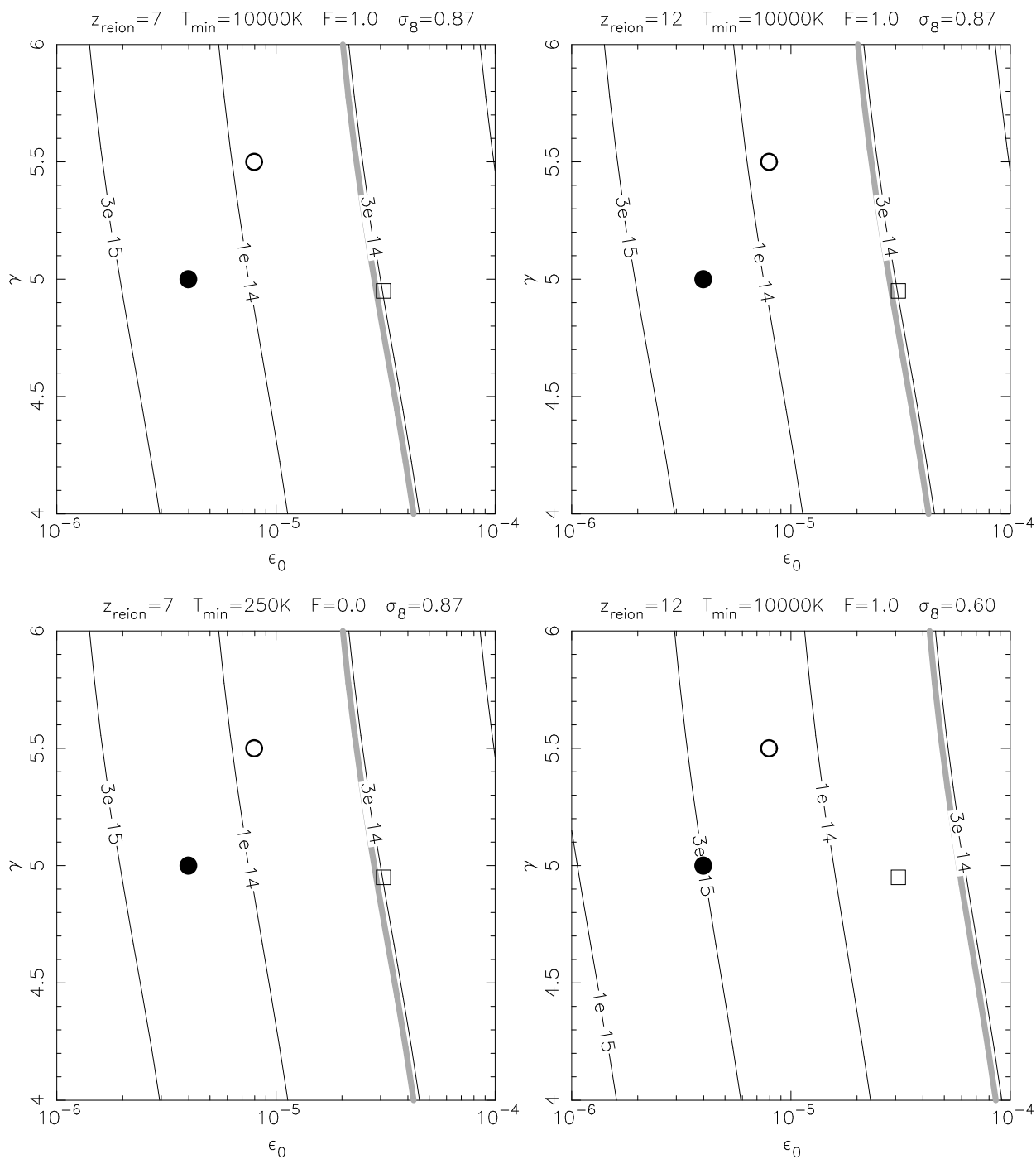}
\caption{\label{fig6} {\footnotesize Contours of the value of the
characteristic strain spectrum at $f_{\rm c}=1.9\times10^{-9}$Hz as a
function of $\epsilon_0$ and $\gamma$. The thick line shows the
current upper limit from pulsar timing experiments. The dot shows the
position of the parameter pair $\epsilon_0=10^{-5.4}$, $\gamma=5$. The
open circle the values found by Ferrarese~(2002) assuming that the
circular velocity equals the virial velocity
($\gamma=5.46,\epsilon_{0}=10^{-5.06}$), while the square shows the
observed values ($\gamma=4.95,\epsilon_{0}=10^{-4.54}$) assuming an
NFW profile (Bullock et al.~2001).  The panels show cases I-IV whose
primary parameters are listed at the top of each plot.}}
\end{figure*}

The models considered in this paper predict 
characteristic strain spectra with values at $f_{\rm
c}=1.9\times10^{-9}$Hz that are close to the current limits from pulsar
timing experiments (Kaspi et al.~1994; Thorsett \& Dewey~1996 and
Lommen~2002).  So far we have concentrated on the spectrum
and event rate for $\epsilon_0=10^{-5.4}$ and $\gamma=5$, which are
the values favored by simple feedback arguments and the theoretical
model for the quasar luminosity function described in Wyithe \&
Loeb~(2002). However variation of $\epsilon_0$ and $\gamma$ has a
significant effect on the GW signal.  Since $h_{\rm spec}$ is
dominated by the mergers of massive ($\ga10^{12}M_\odot$) galaxies at
low redshifts ($z\la2$) in which supermassive BHs are thought
to be ubiquitous, it should be quite insensitive to the details of the
uncertain evolution in the BH mass function at higher
redshifts. The amplitude of $h_{\rm spec}(1.9\times10^{-9}{\rm Hz})$
therefore provides robust constraints on combinations of 
the parameters $\epsilon_{\rm mrg}$, $\gamma$ and $\epsilon_0$.

In Figure~\ref{fig6} we show contours of $h_{\rm
spec}(1.9\times10^{-9}{\rm Hz})$ as a function of $\epsilon_0$ and
$\gamma$. The thick grey line shows the contour at the current limit
of $h_{\rm spec}<3\times10^{-14}$. The dot shows the values
$\epsilon_0=10^{-5.4}$ and $\gamma=5$. As mentioned the value is close
to the observational limits.  The open circle shows the values found
by Ferrarese~(2002) assuming the circular velocity equals the
virial velocity ($\gamma=5.46,\epsilon_{0}=10^{-5.06}$), which is the
assumption implicit within our formalism. The square shows the
corresponding observed values ($\gamma=4.95,\epsilon_{0}=10^{-4.54}$)
obtained assuming a Navarro, Frenk \& White~(1997) profile (Bullock et
al.~2001).  We see that if coalescence is prompt for all
binaries, then the pulsar measurements already exclude parameter
space close to the currently favorable values. 

Adoption of the values determined by Ferrarese~(2002) for the BH --
halo mass relation implies either that the background will soon be
detected, or that the efficiency for BH binary coalescence at low redshift is
$\epsilon_{\rm mrg}<1$. Yu~(2002), investigated the evolution of
hypothetical BH binaries in realistic models of nearby early-type
galaxies. She found that BH binaries are likely to survive for longer
than a Hubble time in spherical or weakly flattened/triaxial
galaxies. No evidence has yet been found for BH binaries in the
centers of the galaxies studied. The upper limit of the range of
semimajor axes of surviving binary black holes as determined by
Yu~(2002) is close to the HST resolution for typical nearby galaxies
(i.e., galaxies at Virgo). Thus, the absence of evidence for double
nuclei in nearby galactic centers does not necessarily mean that they
have no BH binaries (Q.~Yu, private communication).  Further
improvement by an order of magnitude in the upper limits on the GW
background will imply that $\epsilon_{\rm mrg}<1$ and that BH binaries
may survive in galaxies rather than coalesce. Yu~(2002) also computed
the periods of surviving BH binaries. Of particular importance to this
study is that nearly all of the surviving BH binaries would have
periods longer than 20 yrs, and would therefore not contribute to
$h_{\rm spec}(1.9\times10^{-9}Hz)$.

\section{Conclusions}
\label{conc}

We have computed the characteristic strain spectrum for GWs, as well
as the number of events detectable by the planned LISA observatory due
to coalescing massive BH binaries. Rather than adopt a
phenomenological approach which is limited by the redshift horizon of
current observations, we calculated the expected merger history of
galaxies out to arbitrarily high redshifts under different
assumptions.  Our model uses the observed relation (Ferrarese 2002)
between the circular velocity of galaxies and their central BH mass,
and assumes that massive BHs form in galaxies into which gas can
condense and cool.  An important ingredient in the prescription is the
observation that the slope and normalization of the relation holds out
to high redshift (Shields et al. 2002).  The underlying principles of
this approach have been shown to reproduce current data on the
luminosity function of quasars (Wyithe \& Loeb 2002; Volonteri, Haardt
\& Madau~2002).

We find that if BH formation is ongoing in galaxies with masses above
the cooling threshhold, then reionization produces a sharp drop in the
number counts of LISA sources as a function of source redshift (see
Fig. \ref{fig3}).  In the instance that all binary BHs proceed to 
coalescence our models generically predict that hundreds of events 
with strains $h_c>10^{-22}$ will be detectable per year by LISA.
The strongest sources have short
durations and will not suffer from a confusion limit.  Interestingly,
most of these sources originate from redshifts $z>7$.

In the nHz regime accessible to pulsar timing experiments, we find
that the amplitude of the characteristic strain spectrum (related to
the power in the gravitational wave background) is close to current
limits assuming values (favored by Wyithe \& Loeb 2002) of $\gamma=5$
and $\epsilon_0=10^{-5.4}$ for the slope and normalization relating
the BH to host galactic halo mass (see equation~\ref{bh}).  Based on
local observational data on early type galaxies, Ferrarese~(2002)
finds $\gamma=5.46$ and $\epsilon_{0}=10^{-5.06}$ (where the circular
velocity is assumed to equal the virial velocity), or $\gamma=4.95$
and $\epsilon_{0}=10^{-4.54}$ (where the halo density is assumed to
follow an NFW profile). Calculations of the characteristic strain spectrum
assuming these parameter sets are comparable to current limits, 
and already rule out only slightly different combinations
of $\gamma$ and $\epsilon_0$ if all binary black holes proceed to
coalescence. Alternatively, future improvements in the observational
limits will allow limits to be placed on the fraction of black hole
binaries that proceed to coalescence from calculations of the strain
spectrum that use measured values of $\gamma$ and $\epsilon_0$.

We find that the amplitude of the characteristic spectrum of strain is
dominated by the mergers of massive galaxies ($M\ga10^{12}M_\odot$) at
low redshift ($z\la2$). As a result, $h_{\rm spec}$ is quite sensitive
to the value of $\sigma_8$. Since BHs are ubiquitous in the local
universe, detection of the background either by future pulsar timing
measurements or by the LISA satellite will determine the efficiency
with which binary BHs coalesce.

Observations of GWs are likely to provide insight into the formation
history of massive BHs.  LISA will open a new window to studies of the
high redshift universe.  The number counts of gravitational
wave sources will provide an unprecedented probe of the reionization
epoch, and constrain the merger rate of galaxies and the growth
history of black holes at early epochs that are not yet probed by
conventional astronomical telescopes. Pulsar timing arrays are
suitable to measure the background arising from BH binaries at lower
redshifts.

\acknowledgements 

We thank Andrew Melatos for many helpful discussions, and Scott Hughes
and Cathy Trott for comments on the manuscript.  We also thank Peter
Bender for correcting an earlier version of the Figure 6 and for
helpful comments about LISA. AL acknowledges support from the
Institute for Advanced Study at Princeton, the John Simon Guggenheim
Memorial Fellowship, and NSF grants AST-0071019, AST-0204514.

\end{document}